\documentstyle[12pt]{article} \topmargin -20pt \textheight 640pt
\newcommand{\be}{\begin{equation}} \newcommand{\ee}{\end{equation}}
 
\newcommand{\bea}{\begin{eqnarray}} \newcommand{\eea}{\end{eqnarray}}
\newcommand{\ba}{\begin{array}} \newcommand{\ea}{\end{array}}
\newcommand{\beqa}{\begin{eqnarray}} \newcommand{\eeqa}{\end{eqnarray}}
\newcommand{\lsim}{{\;\raise0.3ex\hbox{$<$\kern-0.75em\raise-1.1ex
\hbox{$\sim$}}\;}}
\newcommand{\gsim}{{\;\raise0.3ex\hbox{$>$\kern-0.75em\raise-1.1ex
\hbox{$\sim$}}\;}}

\newcommand{\cL}{{\cal L}} 
 \newcommand{\h}{{1\over2}}
 
 \newcommand{\Tr}{{\rm Tr}}

\newcommand{\DE}{\Delta} 
\newcommand{\ep}{\epsilon} 
\newcommand{\th}{\theta} 
\newcommand{\lm}{\lambda}

\newcommand{\ssu}{$SU(2)_L\times SU(2)_R\times U(1)_{B-L}\,$}
 
 \newcommand{\matr}{\left( \begin{array}}
\newcommand{\ematr}{\end{array} \right)}

 \newcommand{\ga}{{\gamma}}

\begin{document}

\begin{titlepage}

\mbox{}\vspace*{-1cm}\hspace*{8cm}\makebox[7cm][r]{\large  HU-SEFT R
1994-09} \mbox{}\vspace*{-0cm}\hspace*{9cm}\makebox[7cm][r]{\phantom 
{HU-SEFT R 1993-17}} \vspace*{0.5cm}
 \hspace*{10.9cm} \makebox[4cm]{(hep-ph/9409320)} \vfill
 
\Large
 
\begin{center} {\bf  Vector boson pair production in $e^-e^-$
collisions with polarized beams}
 
\bigskip \normalsize {${\rm P. Helde,}^a\:\:{\rm K. Huitu,}^b\:\: {\rm
J. Maalampi}^c\:\: {\rm and} \:\:  {\rm M. Raidal}^{b}$\\[15pt]$^a${\it
Department of Theoretical Physics, University of Tartu}\\$^b${\it
Research Institute for High Energy Physics, University of
Helsinki}\\$^c${\it Department of Theoretical Physics, University of
Helsinki}}
 
{September 1994}

\bigskip

\vfill
 
\normalsize
 
{\bf Abstract} \end{center} 

\normalsize

The $W$-boson pair production  in $e^-e^-$ collisions with polarized
beams is investigated. The helicity amplitudes are derived for general
couplings and the conditions for a good high-energy behaviour of the
cross-section are given. The results are applied to the heavy vector
boson production in the context of the left-right symmetric model. The
Ward identities and the equivalence theorem are also discussed.

\normalsize
 
\end{titlepage}

\section{Introduction}

While colliding  electrons against positrons will be the main operation
mode of the next generation linear collider (NLC), also the
electron-electron option is technically viable \cite{Paula}. Realizing
this possibility would be well motivated from the physics point of
view. The $e^-e^-$ collisions have been so far experimentally explored
only at very modest energies, where the M\o ller scattering  dominates
the total cross section. At high energies many  other processes become
possible providing
 a probe into  interesting  new features of the electroweak
interactions \cite{rev}. 

The $e^-e^-$ mode is particularly ideal for testing the possible lepton
number violation, since the initial state carries, in contrast with the
$e^+e^-$ collisions, a non-vanishing lepton number (see e.g.
\cite{early,MPVnp,Fram}). It is also suitable for search of
supersymmetric extensions of the electroweak models
\cite{Keung,Ruckl,HMR}. The possibility provided by  linear colliders
of having polarized beams offers effective methods for such studies.
From the experimental point of view the electron-electron collision is
an excellent place to study new phenomena, because the Standard Model
(SM) background is low due to non-existence of dominating s-channel
resonancies.

One  interesting  lepton number violating reaction with clear
experimental characteristics is the pair production of massive vector
bosons \cite{early,MPVnp,Minkowski,Tom}:
 
\be e^-_{L,R}e_{L,R}^- \rightarrow W^-W^-, \label{eeWW} \ee

\noindent
 Observation of such a process would be an inevitable signal  of new
physics, since in the Standard Model (SM) the lepton number is
conserved and the reaction (\ref{eeWW}) is absolutely forbidden (the SM
reaction  $e^-e^-\to W^-W^- \nu_e\nu_e$ mimicing (\ref{eeWW}) can be
effectively eliminated by imposing suitable kinematical cuts
\cite{Cuypers}). In a wide class of  other models  beyond the SM the
reaction may occur. For example, in the \ssu left-right symmetric
theory \cite{pati} it is possible   due to the lepton number violating
Yukawa interactions and the existence of Majorana neutrinos
\cite{early,MPVnp}. In  that model the  vector bosons may be ordinary
charged weak bosons, $W_1$, or the heavier right-handed weak bosons,
$W_2$. Depending on the assumptions made on some model parameters the
cross section for the $W_1$ pair production may according to 
\cite{Minkowski} be tens of fb's at $\sqrt{s}=1$ TeV and some fb's at
$\sqrt{s}=0.5$ TeV. The present lower limit for the mass of $W_2$ is
$M_{W_2}\gsim 600$ GeV \cite{WRmass} so that the reaction (\ref{eeWW})
might be kinematically allowed also for the $W_2W_2$ final state, at
least in the final phase of the NLC. Soon above the threshold the cross
section of this reaction may reach the level of pb \cite{MPVpl}.

 The reaction (\ref{eeWW}) can  proceed at tree level through a
 neutral fermion exchange in t- and u-channels (see Fig. 1). As the
lepton number is broken in the process, the neutral  fermions exchanged
have to be self-conjugated  particles. Hence the reaction (\ref{eeWW})
would reveal the nature of neutrinos by making difference between the
cases where neutrinos are Majorana particles and Dirac particles.
Nevertheless, the amplitudes of the  t and u-channels alone would
violate unitarity at high energies. In order to  make the total cross
section  consistent with  unitarity also a lepton number violating
doubly  charged boson exchange in s-channel (see Fig. 1)  has to occur
and should be taken into account.

In a gauge theory the good high-energy behaviour is guaranteed by the
gauge structure and  both the t, u and s-channel processes are present
automatically.  So is the case, e.g., in the left-right symmetric
model, where the doubly charged boson ($\Delta^{--}$) is a spin-0
particle   (and not a vector particle,  dilepton, which appears in
another class of models \cite{Fram}).  The $\Delta^{--}$ is a member of
a right-handed Higgs triplet, which plays a central role in the model
by driving the spontaneous breaking of the left-right symmetry and
thereby giving a large mass to the exotic particles predicted by the
model. The triplet Higgses also give rise to the see-saw mechanism for
neutrino masses \cite{seesaw}, as a result of which there are two
self-conjugate neutral fermions (neglecting the possible interfamily
mixing) contributing to (\ref{eeWW}) in t- and u-channels, a light and
a heavy Majorana neutrino.

In this paper we shall investigate the reaction (\ref{eeWW}) for
polarized beams and with final state polarization measurements. We
derive helicity amplitudes assuming a general form for the relevant
couplings, and  derive   conditions between the s- and t- and u-channel
couplings which should be satisfied in order to guarantee a good high
energy behaviour of the interaction. We also  study how small
deviations from this condition would manifest themselves in measurable
quantities. Such deviations might reflect the possibility that a  gauge
theory, such as the left-right theory, is just an effective theory to
be replaced at short distances by a more fundamental one.

The plan of the paper is the following. In Section 2 we derive the
helicity amplitudes and discuss the high energy behaviour of the cross
section.   In Section 3 we present numerical results.  In Section 4 we
consider the Ward identities and the equivalence theorem. Summary is
given in Section 5.

\section{Helicity amplitudes}

In a gauge theory the longitudinal components of  gauge bosons play a
special role since they exist as a consequence of the Higgs mechanism.
In the original Lagrangian they corres\-pond to Gold\-stone fields, the
scalar fields associated with the spontaneous breaking of the gauge
symmetry. The interactions of the longitudinal gauge bosons thus reflect
the properties of the Higgs scalars responsible for the symmetry
breaking. It is therefore  useful to investigate the reaction
(\ref{eeWW}) by taking into account the polarization of the particles. 
In the following we give the helicity amplitudes of (\ref{eeWW}) to
serve  this purpose.

Let us start by defining the relevant couplings without devoting
ourselves to any particular model. We assume  the interaction needed
for the reaction to proceed through the neutrino exchange in t- and
u-channels to have the form
 
\be \cL^{cc}=\frac g{2\sqrt{2}} \bar\nu \gamma_\mu  (A_1+B_1\gamma_5) e
W^{+\mu} +h.c., \ee
 \noindent where $g$ is a coupling constant and $A_1$ and $B_1$ are 
parameters.  This has the same form as the weak boson - neutrino
coupling  in SM, except that   it allows for a more general
vector-axial-vector structure. For the s-channel process we have to
specify the $W^-W^-\Delta^{++}$ and  $e^-e^-\Delta^{++}$ interactions. 
The interaction between $\Delta^{++}$ and the $W$'s we assume to have 
the form
 
\be \cL_{WW\DE } = \frac 1{ \sqrt{2}} g^2 v W^-W^-\Delta^{++} + h.c.
\label{WWD}\ee
 
\noindent The $v$ can be taken as a free mass parameter making this
interaction independent of the $\nu e^-W^+$ coupling. 

 The lepton number violating Yukawa interaction which determines the
coupling between $\Delta$ and the electrons is taken to be 
 
\be \cL_{l'l\DE}= - \frac 1{\sqrt{2}} h e^T C (A_2+B_2\gamma_5 )
\DE^{++}e +h.c., \label{Yukawa}\ee

\noindent where $h$ is the Yukawa coupling constant and $A_2$ and $B_2$
are free parameters.

 As mentioned, using polarized beams and detecting the final state
polarizations may give valuable insight into the process (\ref{eeWW}).
One issue of interest in such a study would be   the interactions of
the longitudinal components of the vector bosons. In reality the
polarization is never perfect and when calculating cross sections one
needs to apply the  density matrix containing the transversely and
longitudinally polarized components of the beams.

We now write down the helicity amplitudes for the process (\ref{eeWW}).
We will neglect the electron mass, which has the consequence that the
polarized electrons coincide with the left-  and right-handed chirality
states denoted by  $\lm = -\h$ and $\lm = +\h$, respectively. The
electron momenta are assumed to be along the $z$-axis, $p_i =(p,0,0,\pm
p)$ ($i=1,2$). The $W$ momenta are given by  $k_i^\mu=(k^0,\pm k\sin\th
,0,\pm k \cos\th )$ and the polarization vectors of the $W$'s are 

\bea \ep_{\tau =\pm 1}(k_i)&=& \frac 1{\sqrt{2}}  (0,\mp\tau\cos\th,
-i,\pm\tau\sin\th ),\\ \ep_{\tau =0}(k_i) &=& \pm \frac 1{M_W} (\pm
k,k^0\sin\th,0,k^0\cos\th ). \eea
 
\noindent Using the notation above, the helicity amplitudes are given by
 
\bea F_{\lm\lm ' \tau\tau ' } &=& \frac{i2hg^2 v}{s-M_\DE^2}
 \bar v(p',\lm ') (A_2+B_2\gamma_5) u(p,\lm ) \ep^*_\mu
(k,\tau)\ep^{\mu *}(k',\tau ' )  \nonumber \\ && -\frac{i g^2}{8}
 \bar v(p',\lm ') C(i\Gamma_{\mu_2})^TC^{-1}
 \frac{1}{(p\!\!\!/ -k\!\!\!/) -m_\nu )} i\Gamma_{\mu_1} u(p,\lm )
\nonumber \\ && \times \ep^{\mu_1 *} (k,\tau) \ep^{\mu_2 *}(k',\tau ' )
+ (k\leftrightarrow k' ), \eea
 
\noindent where $\Gamma_\mu = ( g/{2\sqrt{2}}) \ga_{\mu}
(A_1+B_1\gamma_5)$. The first two indices in $F_{\lm\lm ' \tau\tau ' }$
denote the electron helicities and the last two  the gauge boson
helicities. More explicitly, the amplitudes read ($\lm =\pm\h $  and
$\tau =\pm 1$, longitudinal $W$'s are denoted by 0):
 
\bea F_{\lm -\lm\tau\tau} &=&
-i\frac{g^2\lm\sqrt{s}}{4\sqrt{s-4M_W^2}}(A_1^2-B_1^2) \sin\th \left\{
\frac{-t-M_W^2}{t-m_\nu ^2} +  \frac{-u-M_W^2}{u-m_\nu ^2}
\right\},\nonumber \\ F_{\lm \lm\tau\tau} &=& -i\frac{g^2\sqrt{s} m_\nu
}8 (A_1+2\lm B_1)^2  \nonumber \\ && \times \left\{ -\frac
{1+2\lm\tau\cos\th}{t-m_\nu ^2}   -\frac {1-2\lm\tau\cos\th}{u-m_\nu ^2}
\right\}  + i \frac{ 2 \sqrt{s} g^2 v h }{ (s-M_\DE ^2)} (A_2+2\lm
B_2),\nonumber \\ F_{\lm -\lm\tau -\tau} &=& -i\frac 1{16}{g^2s\tau
\sin\th} (A_1^2-B_1^2)(1+2\lm\tau\sin\th) \left\{\frac 1{t-m_\nu^2}
+\frac 1{u-m_\nu^2} \right\},\nonumber \\ F_{\lm \lm\tau -\tau} &=&
0,\nonumber \\ F_{\lm -\lm\tau 0} &=& -i \frac{g^2\sqrt{s}
(A_1^2-B_1^2)}{8\sqrt{2}M_W}  (1+2\lm\tau\cos\th)  \nonumber \\ &&
\times \left\{ \left[ -2M_W^2\lm\tau +\sqrt{\frac s{s-4M_W^2}}(t+M_W^2)
\right] \frac 1{t-m_\nu ^2} \right.   \nonumber \\ &&+ \left. \left[
-2M_W^2\lm\tau -\sqrt{\frac s{s-4M_W^2}}(u+M_W^2) \right] \frac
1{u-m_\nu ^2} \right\},\nonumber \\ F_{\lm -\lm 0\tau } &=& -i
\frac{g^2\sqrt{s} (A_1^2-B_1^2)}{8\sqrt{2}M_W}  (1-2\lm\tau\cos\th)
\nonumber \\ && \times \left\{ \left[ 2M_W^2\lm\tau +\sqrt{\frac
s{s-4M_W^2}}(t+M_W^2) \right] \frac 1{t-m_\nu ^2} \right. \nonumber \\
&&+ \left.\left[ 2M_W^2\lm\tau -\sqrt{\frac s{s-4M_W^2}}(u+M_W^2)
\right] \frac1{u-m_\nu ^2}  \right\},\nonumber \\ F_{\lm \lm \tau 0}
&=& -i \frac{g^2\sqrt{s} m_\nu\tau}{16 \sqrt{2} M_W} (A_1+2\lm B_1)^2
(-\sqrt{s-4M_W^2} -2\lm\tau\sqrt{s} ) \sin \th \nonumber \\ && \times
\left( \frac 1{t-m_\nu ^2} - \frac 1{u-m_\nu ^2} \right),\nonumber \\
F_{\lm \lm 0\tau } &=& -i \frac{g^2\sqrt{s} m_\nu\tau}{16 \sqrt{2} M_W}
(A_1+2\lm B_1)^2 (\sqrt{s-4M_W^2} +2\lm\tau\sqrt{s} ) \sin \th
\nonumber \\ && \times \left( \frac 1{t-m_\nu ^2} - \frac 1{u-m_\nu ^2}
\right),\nonumber \\ F_{\lm -\lm 00 } &=& -i \frac{g^2\sin\th
(A_1^2-B_1^2)}{16M_W^2} \sqrt{\frac s{s-4M_W^2}} \left\{ \frac
{-st-4M_W^4}{t-m_\nu ^2} +
 \frac {su+4M_W^4}{u-m_\nu ^2} \right\},\nonumber \\ F_{\lm \lm 00 }
&=& -i \frac{g^2\sqrt{s} m_\nu}{8M_W^2} (A_1+2\lm B_1)^2 \left\{
\frac{-t}{t-m_\nu^2 } +  \frac{-u}{u-m_\nu^2 } \right\} \nonumber \\
&&- i \frac{\sqrt{s} g^2 v h }{ (s-M_\DE ^2)M_W^2} (A_2+2\lm B_2) 
(s-2M_W^2). \label{amp} \eea

\noindent The t- and u-channel diagrams are  proportional either to a
factor $(A_1^2 -B_1^2)$ or $(A_1+2\lm B_1)^2$. In the amplitudes of the
latter type, there is a flip in the neutrino helicity, which is
indicated by the proportionality to the neutrino mass. In order to have
interaction vertices which match the helicity flip,  both electrons have
to be of the same chirality. Due to the annihilation into a scalar
particle, the s-channel contribution is non-zero only when the $W$'s
are  similarly polarized and  both electrons have the same chirality.

In case that the $eW$ interaction is purely right-handed ($A_1=B_1$),
the t- and u-channel contributions exist only when  both electrons are
right-handed. For a pure left-handed interaction ($A_1=-B_1$) the
electrons  should correspondingly be left-handed.   The same is true
for the s-channel and the couplings $A_2$ and $B_2$. In the case of
purely chiral couplings the whole process should thus disappear for
suitably chosen beam polarizations. This is of course only an ideal
situation, since in practice the beams will not be 100 \% polarized.

In the helicity amplitudes given above we have not assumed any
interrelations among the couplings or particle masses but have kept
them all as free parameters. In general this would lead to a
contradiction  with unitarity. In order to have a cancellation of  the
unwanted  terms proportional to $s$ the following conditions should be
fulfilled:

\beqa m_{\nu}(A_1+B_1)^2-4hv(A_2+B_2)&=& 0,\nonumber\\
m_{\nu}(A_1-B_1)^2-4hv(A_2-B_2)&=& 0. \eeqa

\noindent This facilitates our previous statement that both the t, u
and the s channel amplitudes are necessary for a good high-energy
behaviour, and on top of that the neutrino mass and the couplings of
the doubly charged scalar should be suitably related to each other.

\section{Numerical results}
 
In this Section we shall present our numerical results. We shall only
consider the reaction $e^-e^-\to W^-_2W^-_2$, where $W^-_2$ is a heavy
gauge boson predicted by the left-right symmetric model. In the
left-right symmetric model the following relation holds:   

\be m_{\nu}=2hv. \label{massrel} \ee Here $v$ is  the vacuum
expectation value of the neutral member of a "right-handed" Higgs
triplet $(\Delta^{--},\Delta^{-},\Delta^{0})$ which is responsible of
the breaking of the $SU(2)_R\times U(1)_{B-L}$ gauge symmetry. It also
determines the mass of the heavy weak boson $W_2$ through
$M_{W_2}=gv/\sqrt{2}$, where $g$ is the gauge coupling constant
associated with the $SU(2)_R$ interactions (we shall assume it to have
the same value as the SM gauge coupling, $g=0.65$).

 Our formulas are also applicable  to the pair production of the
ordinary weak bosons, but the cross section of this process is in
general too small to have any great phenomenological importance at the
energies of  NLC.
 \footnote {It was argued in ref. \cite{Minkowski} that the cross
section of the pair production of the ordinary $W$ bosons could reach
some tens of fb's at $\sqrt{s}=1$ TeV. This estimate was, however,
based on a calculation where only the t- and u-channels were taken into
account. If also the s-channel is included, as unitarity requires, the
cancellations may reduce the cross section substancially.}
  In all plots to be presented we assume it to have the mass
$M_{W_2}=0.5$ TeV. For the mass of the doubly charged scalar
$\Delta^{--}$    we shall use different values, at lowest
$M_{\Delta}=0.8$ TeV. Varying this mass affects the total cross section
and in particular the angular distributions. In the left-right
symmetric model, providing that $W_1$ and $W_2$, as well as
$\Delta^{--}$ and its possible left-handed counterpart, do not mix as
we shall assume, all couplings relevant here  are purely right-handed,
i.e. $A_1=B_1=A_2=B_2=1$. For this reason we shall consider the
particular case where the electron beams are right-handedly polarized.

In Fig. 2 we plot the cross sections in the case of fully right-handedly
polarized electron beams corresponding to various $W_2$ polarisation
states as a function of the collision energy $\sqrt s$ assuming for the
neutrino mass the value $m_{\nu}= 1$ TeV. In Fig. 2a we assume the
scalar exchanged  in the s-channel to have the mass $M_{\Delta}=0.8$
TeV, in Fig. 2b $M_{\Delta}=10$ TeV. The dominant contribution in both
cases corresponds to the final state where  both $W$'s are
longitudinally polarized, i.e. to the helicity amplitude
$F_{\frac{1}{2}\; \frac{1}{2}\; 0\; 0}$, and it is at the level of 
picobarns. The relative importance of the s-channel contribution 
compared with the t- and u-channel contributions depends on the masses
of $\Delta$ and $\nu$.  This can be seen by comparing the plots  in
Figs. 2a and 2b, and by inspecting the Figs. 3a and 3b, where the cross
sections are presented as a function of neutrino mass for $\sqrt s=1.5$
TeV. The qualitatively different $\sqrt s$--dependence of the 00 final
state cross section in Figs. 2a and 2b follows from the fact that in
Fig. 2a one is above the $\Delta^{--}$ resonance and in Fig. 2b below
it.

The $M_{\Delta}$ dependence is particularly apparent in angular
distributions. In Fig. 4 we plot the differential cross sections for
different $WW$ polarization states in the collision of two right-handed
electrons (we have taken $M_{\nu}= 1$ TeV  and $M_W=0.5$ TeV. The
results corresponding to $M_{\Delta}=0.8$ TeV are marked with $a$, and
those corresponding to  $M_{\Delta}=10$ TeV are marked with $b$. The
distributions for the +1+1, --1--1 and 00 final state polarizations are
completely opposite in these two cases providing information about the
relative importance of the various amplitudes. The other distributions
do not change with $M_{\Delta}$ because the s-channel amplitude does
not contribute to them.

We have also studied how a deviation from the purely right-handed
interaction would manifest itself. Let us assume that the $e\nu W$
interaction has a small left-handed part
$\epsilon\gamma_{\mu}(1-\gamma_5)$. It can be studied in $e_Re_L$
collisions, which vanish for purely right-handed interactions. In Fig.
5 we have plotted the cross section for $e_Re_L\to
W_{2,{+1}}W_{2,{-1}}$ as a function of the parameter $\epsilon$. Taking
ten events as  discovery limit, one finds that NLC with its anticipated
luminosities of $10^{34}$ cm$^{-2}$ s$^{-1}$ would be sensitive to the
deviations corresponding to $\epsilon\gsim 0.03$.

\section{ Ward identities and equivalence theorem  }

According to the well known equivalence theorem \cite{Gaillard,equivth}
the S-matrix element involving longitudinally polarized vector bosons
is  at high energies  up to a small correction the same as the S-matrix
element obtained by replacing the vector bosons by  the corresponding
unphysical Goldstone bosons. In this Section we will consider this
theorem by investigating the pair production of longitudinally polarized
$W_2$ bosons:

\be e^-e^-\to W_{2,{\rm long}} W_{2,{\rm long}}. \label{longlong}\ee
  
The Goldstone boson corresponding to $W_{2,{\rm long}}$ is  the
predominantly sing\-ly charged member of the Higgs triplet, $\Delta^-$
(assuming no mixing of $W_1$ and $W_2$ ). The s-channel amplitude for
(\ref{longlong}) is hence directly controlled by the triplet Higgs
self-interactions giving a valuable insight to the symmetry breaking
sector of the model.

The Ward identities for the process (\ref{longlong}) take the form
\cite{Gaillard}
 \be \frac{p_1^{\mu_1}p_2^{\mu_2}}{M_{W_2}^2} 
S_{\mu_1\mu_2}(p_1,p_2)-S_{4 4}(p_1,p_2)+
i\frac{p_1^{\mu_1}}{M_{W_2}}S_{\mu_1 4}(p_1,p_2)+
i\frac{p_2^{\mu_2}}{M_{W_2}}S_{4 \mu_2 }(p_1,p_2)=0, \label{ward1} \ee

\be i\epsilon^{\mu_1}_{(L)}(p_1)\frac{p^{\mu_2}}{M_W} 
S_{\mu_1\mu_2}(p_1,p_2)-\epsilon^{\mu_1}_{(L)}(p_1)S_{\mu_1
4}(p_1,p_2)=0. \label{ward2} \ee     Here $p_1^{\mu_1}$ and
$p_2^{\mu_2} $ are the  four-momentas of the 
 right handed  vector or Goldstone bosons.   The notation is such that
$ \epsilon_1^{\mu_1}\epsilon_2^{\mu_2} S_{\mu_1\mu_2}(p_1,p_2)$ is the
S-matrix element of the process  $ e^-e^-\rightarrow W^-W^-$, $
\epsilon_1^{\mu_1} S_{\mu_1 4}(p_1,p_2)$ is the S-matrix element of the
process  $ e^-e^-\rightarrow W^-\Delta^-$ and  $S_{44}(p_1,p_2)$ is the
S-matrix element of the process  $ e^-e^-\rightarrow \Delta^-\Delta^-$. 

 Expressing the longitudinal polarization vector of a vector boson with
momentum $ p$ as \be
\epsilon^{\mu}_{(L)}(p)=\frac{p^{\mu}}{M_W}+v^{\mu}(p) , \ee  where $
v^{\mu}(p)$ is a four-vector with the components of the order $
M_W/E$,  one can,  with help of the equation (\ref{ward2}), cast the 
equation (\ref{ward1}) to the form \be S_{4
4}(p_1,p_2)=-\epsilon^{\mu_1}_{(L)}(p_1)\epsilon^{\mu_2}_{(L)}(p_2)
S_{\mu_1\mu_2}(p_1,p_2)+{\cal O}(\frac{M_{W_2}}{E}).  \label{eqth} \ee
This is the equivalence theorem for our case.

The  processes involving Goldstone bosons relevant  for us  are \be
 e^-e^-\rightarrow \Delta^-\Delta^- \label{dd} \ee and \be
 e^-e^-\rightarrow W^-\Delta^-. \label{dw} \ee The Feynman diagrams of
these reactions are given in Figs. 6 and 7.  The relevant Feynman rules
follow from the kinetic  Lagrangian, \be {\cal
L}=(D_{\mu}\Delta_R)^{\dagger}(D^{\mu}\Delta_R), \label{kin} \ee from
the scalar potential \be V=-\mu^2\Tr
(\Delta_R\Delta_R^{\dagger})+\rho_1(\Tr(\Delta_R\Delta_R^{\dagger}))^2+
  \rho_2\Tr (\Delta_R\Delta_R)\Tr
(\Delta_R^{\dagger}\Delta_R^{\dagger}), \label{pot} \ee and from the
Yukawa coupling \be \cL_{Y}= - \frac 1{\sqrt{2}} h \psi_l^T C
(A_2+B_2\gamma_5 ) \DE \psi_l +h.c., \label{Yukawa2}\ee where
$\psi_l=(\nu,l^-)$.

  A straightforward calculation gives for the
 helicity amplitude of the process (\ref{dd}) the expression: \be
F_{\lambda\lambda}^{\Delta\Delta}  =  -i4h^2m_{\nu}\sqrt{s}
       \left\{ \frac{1}{(t-m^2_{\nu})}+\frac{1}{(u-m^2_{\nu})} \right\} 
  +i\frac{32\rho_2\sqrt{s}hv}{(s-M^2_{\Delta})}. \label{hadd} \ee
According to the equivalence theorem (\ref{eqth}) this should
approach   in the limit $ s\rightarrow\infty$   the  helicity amplitude
$F_{\lambda\lambda 00}$ with the reversed sign. Indeed, replacing $
h,v$ and $ \rho_2$ by their left-right model expressions from the
formulas $ M_W^2=g^2v^2/2$, $ M^2_{\Delta}=8\rho_2v^2$ and $
m_{\nu}=2hv$ we see that this is the case. However, it is worth of 
noticing   that the equivalence theorem is applicable for
approximations only if not just the mass of final state vector bosons
but  also that of the virtual doubly charged scalar is small compared
with the collision energy.

Let us now turn to the Ward identity (\ref{ward1}).  In order to make
use of this identity we have to calculate quantities  $
({p_1^{\mu_1}p_2^{\mu_2}}/{M_W}^2)  S_{\mu_1\mu_2}(p_1,p_2)$ and $
({p_1^{\mu_1}}/{M_W})S_{\mu_1 4}(p_1,p_2).$ The first of them turns out
to be equal to the helicity amplitude
 $ F_{\lambda\lambda 00}$ given in (\ref{amp}). The second quantity
denoted as $F_{\lambda\lambda}^{\Delta W}$ is given by  \be
F_{\lambda\lambda}^{\Delta W}=-\frac{\sqrt{2}gh\sqrt{s}}{M_W} \left\{
\frac{t}{(t-m^2_{\nu})}+\frac{u}{(u-m^2_{\nu})}-2 
  \frac{(s-M^2_{W})}{(s-M^2_{\Delta})}\right\}. \label{hadw} \ee
Obviously $ F_{\lambda\lambda}^{W \Delta }=F_{\lambda\lambda}^{\Delta
W}$.

The identity (\ref{ward1})
 can be used, at least in princible, for extracting information about 
the triplet Higgs potential  (\ref{pot}).  Discovery of the process $
e^-e^-\rightarrow W_2^-W_2^-$ would enable one to measure the mass of
the gauge boson and thus also the value of the vacuum expectation value
of $\Delta^0$ determined by $ \mu^2$ and $ \rho_1$. By studying other
lepton number violating processes like  $ e^-e^-\rightarrow \mu^-\mu^-$
\cite{Mara}  and  also $ e^-e^+\rightarrow W^-W^+$ \cite{MPVnp}  one
could determine the  gauge and Yukawa coupling constants $ g$ and $ h$,
as well as the masses  $ M_{\Delta}$ and $ m_{\nu}$. Using these
experimental data one can cross check the  model.

In order to get a feeling how sensitive such a test could be, let us
violate the gauge model relation (\ref{massrel}) by writing it as \be
m_{\nu}=2(1+\delta)hv, \ee

\noindent  where $ \delta$ is a parameter. Then the Ward identity
(\ref{ward1}) takes the form
\newpage
 \bea
\frac{g^2}{2(1+\delta)}\left\{\frac{(1-\delta)^2t-m^2_{\nu}}{t-m^2_{\nu}}+
\frac{(1-\delta)^2u-m^2_{\nu}}{u-m^2_{\nu}}\right\} \nonumber \\
-\frac{g^2s}{s-M^2_{\Delta}}+\frac{16\rho_2M^2_{W_2}}{s-M^2_{\Delta}}=0.
\label{rho} \eea Let us emphasize that in this expression all the 
parameters except $ \rho_2$ have their experimentally measured values.
Therefore, together with the kinematical constraint $
s+t+u=2M^2_{W_2}$, the formula (\ref{rho}) determines the parameter $
\rho_2$. 

In Fig. 8 we present the dependence of $\rho_2 $ on  $ M_{\Delta}$ for
the three different values of  the parameter $ \delta$: a) $
\delta=-0.2$, b) $ \delta=0$ and c) $ \delta=0.2$ at the fixed $
\sqrt{s}=1.5$ TeV, $ M_W=0.5$ TeV and $ m_{\nu}=1$ TeV.   The case
$\delta=0$ corresponds to the left-right model and the Ward identity
(\ref{rho}) just reproduces the gauge theory relation between masses, $
 \rho_2={g^2M^2_{\Delta}}/({16M^2_{W_2}}). $ 

\bigskip

\section{Summary}

The next generation linear colliders will provide us a possibility to
study high-energy  electron-electron collisions in any  combinations of
polarization. This will be very useful option in studying new physics 
because of a low background from the Standard Model phenomena.  In this
paper we have studied the reaction $e^-e^-\to W^-W^-$ where $W$ is a
charged vector boson. Starting with arbitrary particle masses and a
quite general form for the couplings involved, we have derived  the
helicity amplitudes of the reaction. We have also given  the conditions
for a good high-energy behaviour of the cross section. We have explored
how one could get information about the underlying theory by using
polarized beams and measurement of the vector boson helicities. 

Our numerical results concern the pair production of heavy vector
bosons, $W_2$, in the framework of the left-right symmetric model. The 
cross section is dominated by the  longitudinally polarized $W_2$'s.
Its value depends on the masses of $W_2$, the heavy Majorana neutrino
and the doubly charged Higgs boson, but for a feasible choice it is on
the level of a few picobarns. As longitudinal components correspond to
the Goldstone bosons, the measurement of this process would give an
insight to the breaking of the left-right symmetry. We also discuss the
Ward identities and the equivalence theorem, which also are useful in
studying the symmetry breaking sector of the left-right model.

\bigskip \noindent{\bf Acknowledgement.} One of the authors (M.R.)
expresses his gratitude to Emil Aaltosen s\"a\"ati\"o, Jenny ja Antti Wihurin
s\"a\"ati\"o and Viro-s\"a\"ati\"o and another one (P.H.) to CIMO for grants.
This work has
been supported by the Academy of Finland

\newpage

\section*{Figure captions}

\medskip

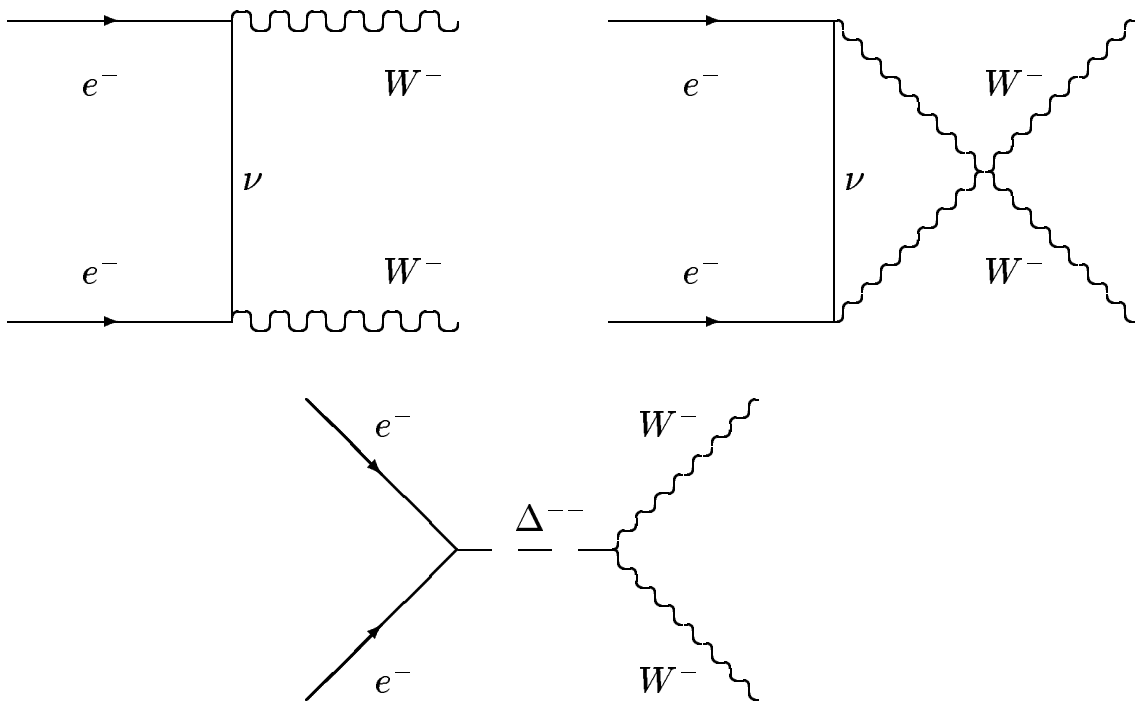
\noindent {\bf Figure 1:}
 Feynman diagrams for the  process $\protect e^-e^-\rightarrow W^-W^-$.

\vspace{0.2cm}

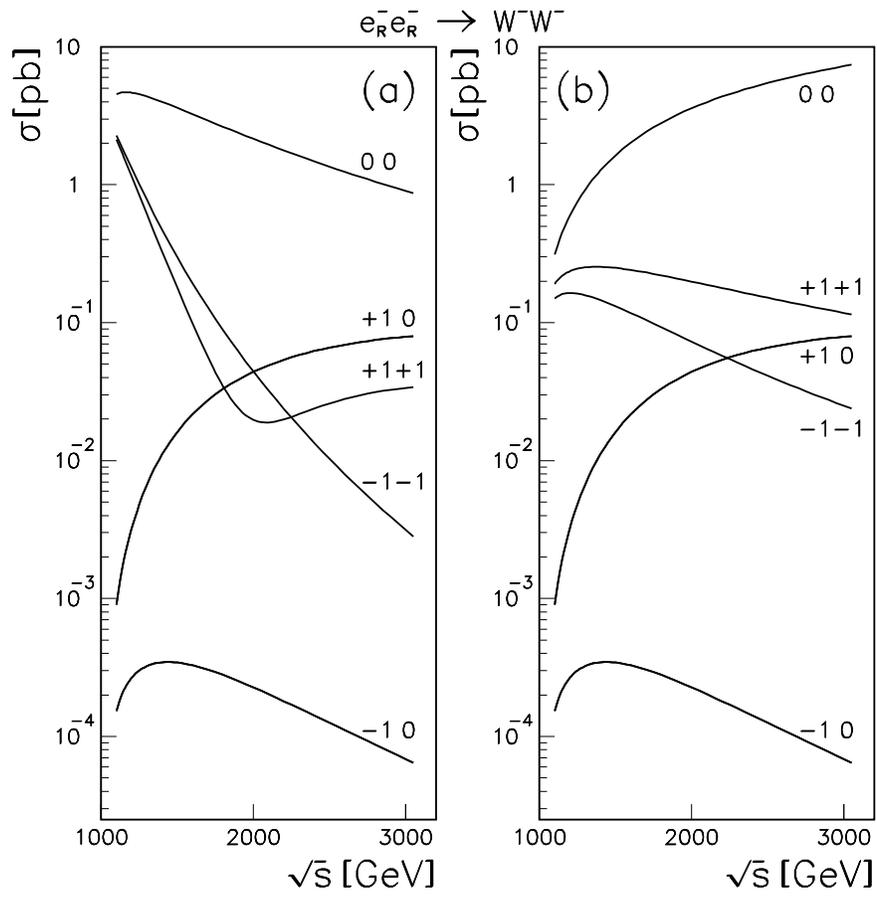
\noindent {\bf Figure 2:} The total cross section of various $\protect
W_2^-W_2^-$
 polarization states for  fully right-handedly polarized beams as a
function of collision energy $\protect \sqrt{s}$.  The mass of
$\protect W_2$ is taken to be $\protect M_W=0.5$ TeV, the mass of heavy
neutrino $\protect m_{\nu}=1$ TeV  and the mass of doubly charged Higgs
boson   a) $\protect M_{\Delta}=0.8$ TeV and   b) $\protect M_{\Delta}=10$
TeV. 
\vspace{0.2cm}

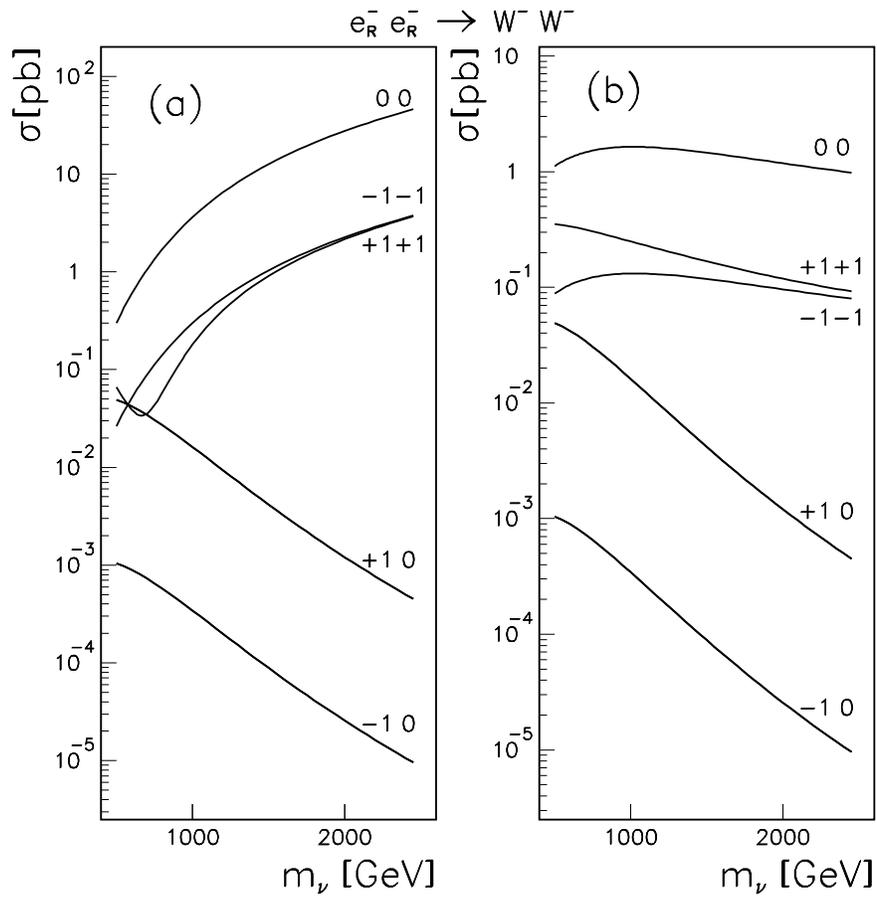
\noindent {\bf Figure 3:} The total cross section of various $\protect
W_2^-W_2^-$
 polarization states for  fully right-handedly polarized beams as a
function of neutrino mass $\protect m_{\nu}$.  The collision energy is
$\protect \sqrt{s}=1.5$ TeV,  the mass of $\protect W_2$ is taken to be
$\protect M_W=0.5$ TeV
 and the mass of doubly charged Higgs boson   a) $\protect M_{\Delta}=0.8$
TeV and  b) $\protect M_{\Delta}=10$ TeV. 

\vspace{0.2cm}

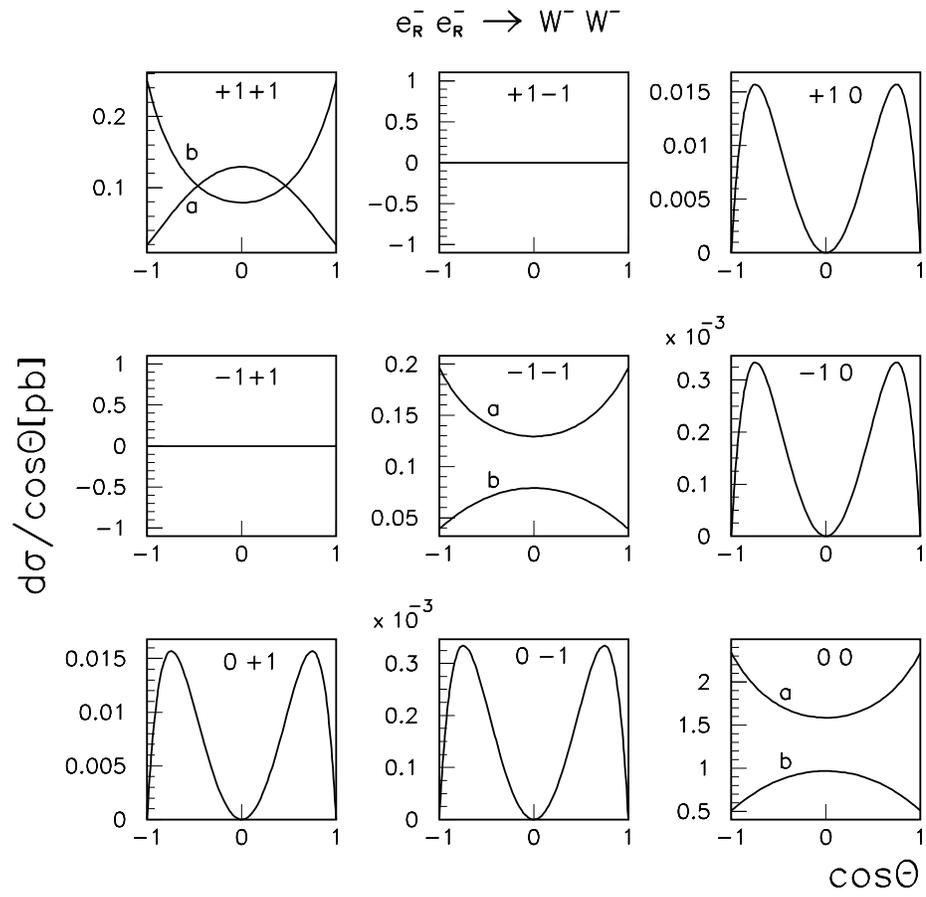
\noindent {\bf Figure 4:} The angular distribution of differential 
cross section of various $\protect W_2^-W_2^-$ polarization states for 
fully right-handedly polarized beams.  The mass of $\protect W_2$ is
taken to be $\protect M_W=0.5$ TeV, the mass of heavy neutrino
$\protect m_{\nu}=1$ TeV  and the mass of doubly charged Higgs boson  a)
$\protect M_{\Delta}=0.8$ TeV  and  b) $\protect M_{\Delta}=10$ TeV. The
off-diagonal distributions do not depend on $\protect M_{\Delta}$.

\vspace{0.2cm}

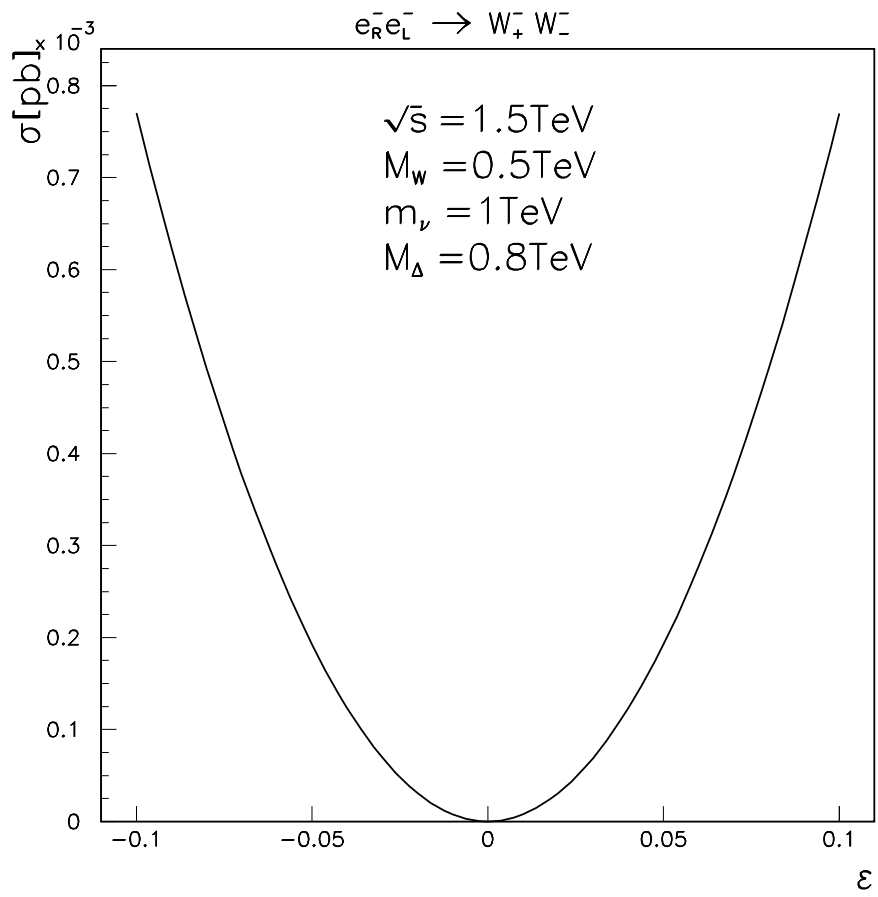
\noindent {\bf Figure 5:} The total cross section of the process
$\protect e_R^-e_L^-\rightarrow W_{+1}^-W_{-1}^- $  as a function of
parameter  $\protect\epsilon$, the fraction of the V-A coupling. 

\vspace{0.2cm}

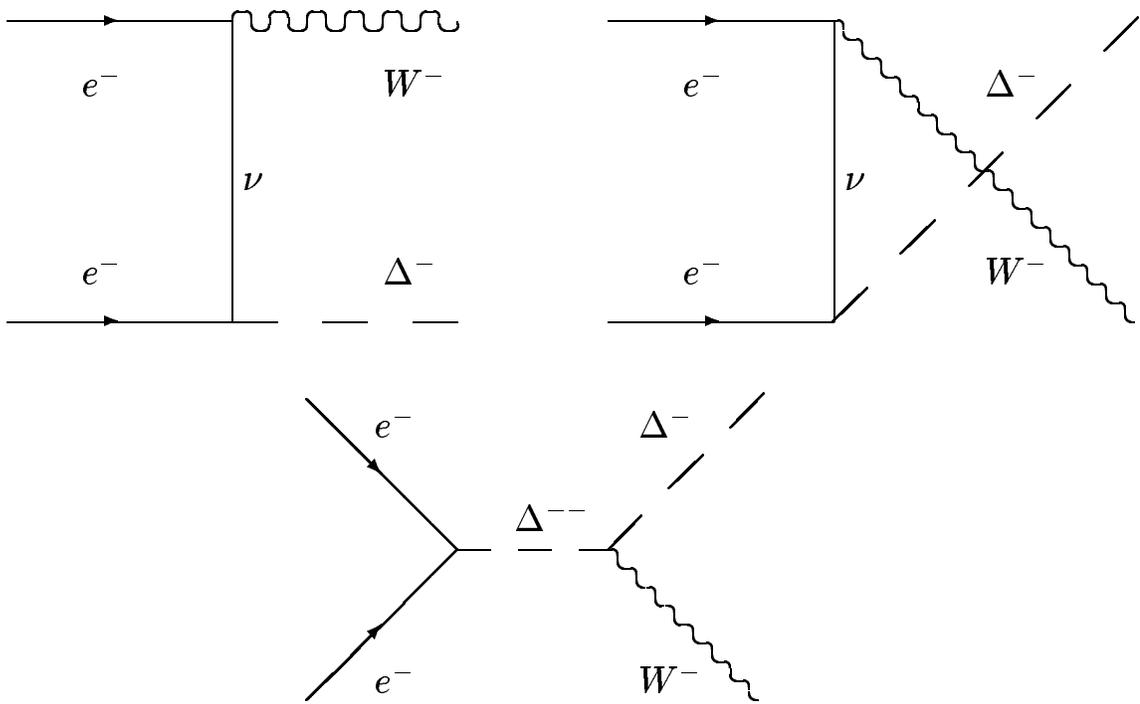
\noindent {\bf Figure 6:} Feynman diagrams for  the  process $\protect
e^-e^-\rightarrow \Delta^-W^-$.

\vspace{0.2cm}

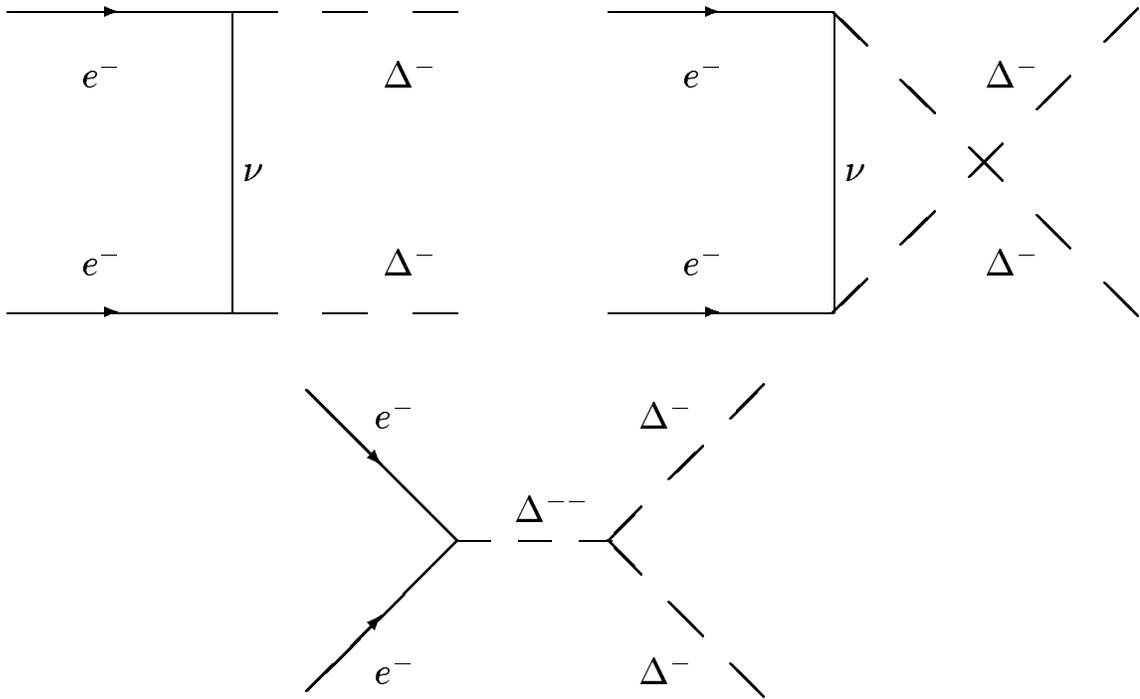
\noindent {\bf Figure 7:} Feynman diagrams for the   process $\protect
e^-e^-\rightarrow \Delta^-\Delta^-$.

\vspace{0.2cm}

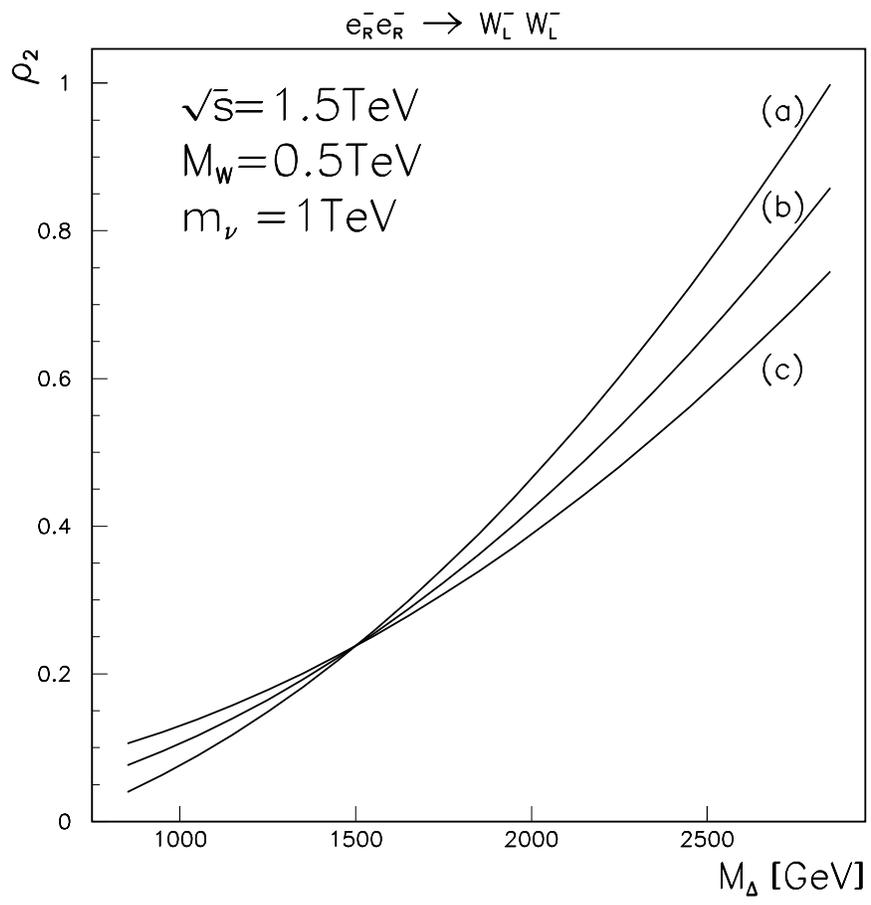
\noindent {\bf Figure 8:} The dependence of  the parameter
$\protect\rho_2$  on the mass of doubly charged  Higgs boson for the
three values of parameter  $\protect\delta$: a) $\protect\delta=-0.2$,
b) $\protect\delta=0$ and c) $\protect\delta=0.2$. Here
$\protect\delta$  measures the deviation of the heavy neutrino mass
form its left-right model value.


\begin{thebibliography}{99}

\bibitem{Paula}  See e.g., {\em Proc. of the  Workshop on physics and
experiments with linear colliders} (Saariselk\"a, Finland, September
1991), eds. R. Orava, P. Eerola and M. Nordberg (World Scientific,
1992), and {\em Proc. of the  Workshop on physics and experiments with
linear colliders} (Waikoloa, Hawaii, April 1993), eds. F.A. Harris,
S.L. Olsen, S. Pakvasa and X.Tata (World Scientific, 1993).

\bibitem{rev} See e.g., F. Cuypers, K. Kolodziej, and R. R\"uckl,
preprint MPI-PhT/94-33, and references therein.


\bibitem{early}  T. Rizzo, Phys. Lett. B116 (1982) 23;\\
 D. London, G. Belanger and J.N. Ng, Phys. Lett. B188 (1987) 155.


\bibitem{MPVnp} J. Maalampi, A. Pietil\"a and J. Vuori, Nucl. Phys. B
381 (1992) 544.

\bibitem{Fram} P.H. Frampton and B.-H. Lee, Phys. Rev. Lett. 64 (1990)
619; P.H. Frampton, Mod. Phys. Lett. A7 (1992) 559 and 2017.
\bibitem{Keung} W.-Y. Keung and L. Littenberg, Phys. Rev. D28 (1983)
1067.


\bibitem{Ruckl} F.Cuypers, G. van Oldenborgh, R. R\"uckl,
    Nucl. Phys. B409 (1993) 128.

\bibitem{HMR} 
 K. Huitu, J. Maalampi, M. Raidal,  Nucl. Phys. B 420 (1994) 449 and  
 Phys. Lett. B 328 (1994) 60.
 
 

\bibitem{Minkowski} C.A. Heusch and P. Minkowski, Nucl. Phys. B416
(1994) 3.

\bibitem{Tom} T. Rizzo, in {\em Proc. of the  Workshop on physics and
experiments with linear colliders} (Waikoloa, Hawaii, April 1993), eds.
F.A. Harris, S.L. Olsen, S. Pakvasa and X.Tata (World Scientific,
1993), p. 520, and 
 preprint SLAC-PUB-6475 (1994).

\bibitem{Cuypers} F. Cuypers, K. Kolodziej, O. Korakianitis and R.
R\"uckl, Phys. Lett. B325 (1994) 243;  V. Barger, J.F. Beacom, K.
Cheung and T. Han, preprint MAD/PH/779 (1994).

\bibitem{pati} J.C. Pati and A. Salam, Phys. Rev. D 10 (1974) 275;\\
R.N. Mohapatra and J. C. Pati, Phys. Rev. D11 (1975) 566, 2558;\\ G.
Senjanovic and R. N. Mohapatra, Phys. Rev. D12 (1975) 1502;\\ R. N.
Mohapatra and R. E. Marshak, Phys. Lett. 91B (1980) 222.
 

\bibitem{WRmass} F. Abe et al., Phys. Rev. Lett. 67 (1991) 2609.

\bibitem{MPVpl}  J. Maalampi, A. Pietil\"a and J. Vuori, Phys. Lett. B
297 (1992) 327.

\bibitem{seesaw} M. Gell-Mann, P. Ramond and R. Slansky, in {it
Supergravity}, eds. P. van Niewenhuizen and D. Z. Freedman (North
Holland 1979);\\ T. Yanagida, in Proceedings of {it Workshop on Unified
Theory and Baryon Number in the Universe}, eds. O. Sawada and A.
Sugamoto (KEK 1979).



\bibitem{Gaillard} M. Chanowitz and M.K. Gaillard, Nucl. Phys. B 261
(1985) 379.

\bibitem{equivth} J.M. Cornwall, D.N. Levin and G. Tiktopoulos, Phys.
Rev. D 10 (1974) 1145; C.E. Vayonakis, Lett. Nuovo Cimento 17 (1976)
383; B.W. Lee, C. Quigg and H. Thacker, Phys. Rev. D 16 (1977) 1519; G.
Gounaris, R. Kögerler and h. Neufeld, Phys. Rev. D 34 (1986) 3257.

\bibitem{Mara} J. Maalampi, A. Pietil\"a and M. Raidal, 
 Phys. Rev. D 48 (1993) 4467.

 \end{thebibliography}
\end{document}